\documentclass[11pt]{article}
\usepackage{amssymb}
\newtheorem{definition}{Definition}
\newtheorem{theorem}{Theorem}
\newcommand{\re}{\mathrm{e}} % Roman e for exponential
\newcommand{\ri}{\mathrm{i}} % Roman i for imaginary number
\newcommand{\rd}{\mathrm{d}} % Roman d for differential
\begin{document}

\title{Accardi complementarity in $\mu$-deformed quantum mechanics}
\author{Claudio de Jes\'{u}s Pita Ruiz Velasco\\Universidad Panamericana\\Mexico City, Mexico\\email: cpita@mx.up.mx \setcounter{footnote}{6}
\and Stephen Bruce Sontz \\Centro de Investigaci\'{o}n en Matem\'{a}ticas, A.C. (CIMAT)\\Guanajuato, Mexico\\email: sontz@cimat.mx}

\date{13 December 2005}

\maketitle

\begin{abstract}
In this note we show that the
momentum and position operators of $\mu$-deformed quantum mechanics for $\mu > 0$
are not Accardi complementary in a sense that we will define.
We conjecture that this is also true if $-1/2 < \mu < 0$.
\end{abstract}

\label{first}

\bigskip

\section[]{Introduction}

\bigskip

We begin by reviewing some basics of $\mu$-deformed quantum mechanics.
This comes from Rosenblum \cite{ros}.
For recent related work see references \cite{as}, \cite{ps} and \cite{pi2}.
We consider a deformation of quantum mechanics depending on
a parameter $\mu > -1/2$, which will be fixed throughout this discussion.
We work in the complex Hilbert space $L^2 ( {\mathbb R}  , m_\mu )$,
where the measure $m_\mu$ for $x \in {\mathbb R}  $ (the real line)
is given by $\rd m_\mu (x) := [ 2^{\mu+1/2} \Gamma (\mu + 1/2) ]^{-1} | x |^{2\mu}
\rd x$.
Here $\rd x$ is Lebesgue measure on ${\mathbb R}$
and $\Gamma$ is the Euler gamma function.
(The normalization constant will be explained later.)
In this Hilbert space we have two unbounded self-adjoint operators: $Q_\mu$,
the $\mu$-deformed position operator, and $P_\mu$, the $\mu$-deformed momentum
operator.
They are defined for $x \in {\mathbb R}  $ and certain elements $\psi \in L^2 (
{\mathbb R}  , m_\mu )$ by
\begin{eqnarray*}
        Q_\mu \psi (x) & := & x \psi(x),  \\
        P_\mu \psi(x) & := & \frac{1}{\ri} \left( \psi^\prime (x) +
                                   \frac{2\mu}{x} ( \psi(x) - \psi(-x) \right).
\end{eqnarray*}
We omit details about
exact domains of definition.
Interest in these operators originates in Wigner \cite{wig}
where equivalent forms of them are used as examples of operators that do not satisfy
the usual canonical commutation relation in spite of the fact that they
do satisfy the equations of motion $ [ H_\mu , Q_\mu ] = P_\mu$ and
$ [ H_\mu , P_\mu ] = -Q_\mu$
for the Hamiltonian
$H_\mu:= \frac{1}{2} ( Q_\mu^2 + P_\mu^2)$.
What does hold is the
$\mu$-deformed canonical commutation relation:
$ \ri [ P_\mu, Q_\mu ] = I + 2 \mu J$, where $I$ is the identity operator
and $J$ is the parity operator $J\psi(x) := \psi(-x)$.

     Many concepts from ordinary analysis also have $\mu$-deformations.
This material also comes form Rosenblum \cite{ros}.
We start with
a $\mu$-deformed factorial function $\gamma_\mu(n)$
defined recursively for integers $n \ge 0$  by $\gamma_\mu(0) := 1$ and
\[
\gamma_\mu(n) := (n + 2 \mu \theta(n)) \gamma_\mu(n-1)
\]
for $n \ge 1$.
Here $\theta$ is the characteristic function of the odd integers.
Using this, we define a $\mu$-deformed exponential function $\exp_\mu(z)$
for $ z \in {\mathbb C}  $ by
\[
 \exp_\mu(z) := \sum_{n=0}^\infty \frac{1}{\gamma_\mu(n)} z^n.
\]
This can be shown to be a holomorphic (entire) function of $z$.
Next, we define a  $\mu$-deformed Fourier transform ${\cal F}_\mu$ by
\[
{\cal F}_\mu \psi (k) := \int_{  {\mathbb R}  } \rd m_\mu(x)
\exp_\mu ( - \ri k x ) \psi(x)
\]
 for $k \in  {\mathbb R}  $ and
$\psi \in L^1 (  {\mathbb R}  , m_\mu )$. In analogy with the
well-known case when $\mu = 0$, this can be shown to define
uniquely a unitary onto transform at the level of $L^2$ spaces,
that is $ {\cal F}_\mu : L^2 (  {\mathbb R}  , m_\mu ) \to L^2 (
{\mathbb R}  , m_\mu )$ is an isomorphism of Hilbert spaces. Given
the formula for ${\cal F}_\mu$, this unitarity condition fixes the
normalization constant in the definition of $m_\mu$.

In \cite{acc} Accardi introduced a definition of complementary observables
in quantum mechanics.
We now generalize that definition to the current context.
We use the usual identification of observables in quantum mechanics
as self-adjoint operators acting in some Hilbert space.

\begin{definition}
We say that two self-adjoint operators $S$ and $T$ acting in $L^2 (  {\mathbb R}  ,
m_\mu )$
are {\em Accardi
complementary} if for any pair of bounded Borel subsets $A$ and $B$ of $  {\mathbb
R}   $ we
have that the operator $ E^S (A) E^T (B) $ is trace class with trace given by:
\[
Tr \left( E^S (A) E^T (B) \right) = m_\mu (A) m_\mu (B).
\]
\end{definition}

Here $E^S$ is the projection-valued measure on $ {\mathbb R}  $
associated with the self-adjoint operator $S$ by the spectral theorem, and
similarly for $E^T$.

So, $E^S (A) E^T (B) $ is clearly a bounded operator acting on $L^2 (  {\mathbb R}
, m_\mu )$.
But whether it is also trace class is another matter.
And, given that it is trace class, it is a further matter to determine if the trace
can be written
as the product of two measures, as indicated.
Accardi's result in \cite{acc} (which is also discussed in detail and proved in
\cite{cas-var})
is that $Q \equiv Q_0 $ and $P \equiv P_0 $ are Accardi complementary.
Accardi also conjectured that this property of $Q$ and $P$ characterized this pair
of operators
acting on $L^2 (  {\mathbb R}  , m_0 )$.
It turns out (see \cite{cas-var}) that this is not so.

\bigskip

\section[]{The Main Result}

\bigskip

We now ask whether the
operators $Q_\mu$ and $P_\mu$ are
Accardi complementary.
To begin this analysis, we will use the following
intertwining relation between these operators given by the
$\mu$-deformed Fourier transform ${\cal F}_\mu$,
which is proved in \cite{ros}: $P_\mu = {\cal F}_\mu^* Q_\mu {\cal F}_\mu$.
This implies the corresponding intertwining relation between their associated
projection valued measures, that is
$E^{P_\mu} (B) = {\cal F}_\mu^* E^{Q_\mu}(B) {\cal F}_\mu $
for every Borel subset $B$ of $\mathbb{R}$.
We wish to calculate the trace of $ E^{Q_\mu}(A) E^{P_\mu}(B) =
E^{Q_\mu}(A) {\cal F}_\mu^* E^{Q_\mu}(B) {\cal F}_\mu$,
where $A$ and $B$ are bounded, Borel subsets of $\mathbb{R}$.
To aid us we define an auxiliary operator
\[
K:= M_e {\cal F}_\mu^* E^{Q_\mu}(B) {\cal F}_\mu M_e : L^2 (
{\mathbb R}  , m_\mu ) \to L^2 (  {\mathbb R}  , m_\mu ),
\]
where $(M_e \psi) (x):= e(x) \psi(x)$ is the multiplication operator
by any $C^\infty$ function of compact support $e: \mathbb{R} \to \mathbb{R}$
satisfying $e(x) = 1$ for all $ x \in A$.
Such a function exists since $A$ is bounded.
Note that $K$ depends on $A$, $B$ and the choice of the function $e$.

We will now calculate the action of $K$ on $\psi \in L^2 (  {\mathbb R}  , m_\mu )$.
We let $\chi_B : \mathbb{R} \to \mathbb{R} $ denote the characteristic function of
$B$.
So for any $ x \in \mathbb{R} $ we have that \pagebreak
\begin{eqnarray*}
   K \psi (x) & = & \left( M_e {\cal F}_\mu^* E^{Q_\mu}(B) {\cal F}_\mu M_e \psi
\right)
\left( x \right) \\
   & = & e(x) \left( {\cal F}_\mu^* E^{Q_\mu}(B) {\cal F}_\mu M_e \psi \right)
\left( x
\right) \\
 & = & e(x) \!\!\int_{ \mathbb{R} } \!\rd m_\mu(k) \exp_\mu ( \ri kx ) \left( E^{Q_\mu}(B)
{\cal F}_\mu M_e \psi \right) \left( k \right) \\
  & = &\! e(x) \!\!\int_{ \mathbb{R} } \!\rd m_\mu(k) \exp_\mu ( \ri kx ) \chi_B(k) \left(
{\cal F}_\mu M_e \psi \right) \left( k \right) \\
  & = \!\!& e(x) \!\!\int_{ \mathbb{R} } \!\rd m_\mu(k) \exp_\mu ( \ri kx ) \chi_B(k) \!\!\int_{
\mathbb{R} } \!\rd m_\mu(y) \exp_\mu ( -\ri ky )
\left(  M_e \psi \right) \left( y \right) \\
  & = & e(x)\!\! \int_{ \mathbb{R} } \rd m_\mu(k) \exp_\mu ( \ri kx ) \chi_B(k) \!\!\int_{
\mathbb{R} } \rd m_\mu(y) \exp_\mu ( -\ri ky )
     e(y) \psi (y) \\
  & = & \int_{ \mathbb{R} } \!\!\rd m_\mu(y)
         \!\!\left[ \int_{ \mathbb{R} } \!\!\rd m_\mu(k) \!\exp_\mu ( \ri kx ) \chi_B(k)
\!\exp_\mu ( -\ri ky ) e(x) e(y)
         \right]
      \!\!\psi (y).
\end{eqnarray*}

This exhibits $K$ as an integral kernel operator with kernel given by
\[
    K(x,y) := e(x) e(y)  \int_{ \mathbb{R} } \rd m_\mu(k) \exp_\mu ( \ri kx )
\chi_B(k) \exp_\mu ( -\ri ky )
\]
for $x,y \in \mathbb{R}$.
(As is conventional, we use the same symbol for the operator and its kernel and let
context indicate
the meaning.)
Clearly, we have that $K$ is $C^\infty$ with compact support in $ \mathbb{R} \times
\mathbb{R} $.
Moreover, on the diagonal we have
\[
          K(x,x) = e(x)^2  \int_{ \mathbb{R} } \rd m_\mu(k) | \exp_\mu ( \ri kx )
|^2 \chi_B(k) \ge 0.
\]

We now can do the central calculation for $A,B$ bounded Borel sets:
\begin{eqnarray*}
Tr \left( E^{Q_\mu}(A) E^{P_\mu}(B) \right) & = &
Tr \left( E^{Q_\mu}(A) {\cal F}_\mu^* E^{Q_\mu}(B) {\cal F}_\mu \right) \\
& = & Tr \left( E^{Q_\mu}(A) {\cal F}_\mu^* E^{Q_\mu}(B) {\cal
F}_\mu  E^{Q_\mu}(A)
\right) \\
& = & Tr \left( E^{Q_\mu}(A)  M_e {\cal F}_\mu^* E^{Q_\mu}(B)
{\cal F}_\mu M_e
E^{Q_\mu}(A)   \right) \\
& = & Tr \left( E^{Q_\mu}(A)  K E^{Q_\mu}(A)   \right) \\
& = & Tr \left( E^{Q_\mu}(A)  K  \right)
= \int_A \rd m_\mu (x) K(x,x) \\
& = & \int_A \rd m_\mu (x) e(x)^2 \int_{ \mathbb{R} } \rd m_\mu
(k) | \exp_\mu ( \ri kx
) |^2 \chi_B (k)  \\
& = & \int_A \rd m_\mu (x)  \int_{ B } \rd m_\mu (k) | \exp_\mu (
\ri kx ) |^2.
\end{eqnarray*}

The step where we evaluated the trace by the (obvious) integral can be justified
using Lemma~1 of \cite{cas-var}, provided that $ 0 \notin A^-$ ($=$ the closure of
$A$) and
$e$ is chosen so that $0 \notin \mathrm{supp}(e)$.
(Take $X = \mathbb{R} \setminus \{ 0 \}$ in \cite{cas-var}, so that $K$ has compact
support
in $ X \times X$ and the density of $m_\mu$ in $X$ is $C^\infty$ and strictly
positive.

Lemma~1 in \cite{cas-var} also asserts that $E^{Q_\mu}(A)  K$ is trace class.)
Of course, we find Accardi's result as the special case $\mu=0$ of this formula,
since then the integrand is identically equal to $1$,
and so the right hand side reduces to $m_0(A) m_0(B)$.
(When $\mu=0$, the technical hypothesis $ 0 \notin A^-$ is not needed.)

We are now ready to state our main result.
\begin{theorem}
Let $A$ and $B$ be bounded Borel subsets of $\mathbb{R}$ with $ 0
\notin A^-$. Then $ E^{Q_\mu}(A) E^{P_\mu}(B) $ is a trace class
operator in $ L^2 (  {\mathbb R}  , m_\mu ) $ for any $\mu > -1/2$
with
\begin{eqnarray}
\label{trace}
0 \le Tr \left( E^{Q_\mu}(A)
E^{P_\mu}(B) \right) = \int_A \rd m_\mu (x)  \int_{ B } \rd m_\mu
(k) | \exp_\mu ( \ri kx ) |^2 < \infty.
\end{eqnarray}

Moreover, if $\mu > 0$ and $m_\mu(A) \ne 0 \ne m_\mu(B)$,
then we have that

\begin{eqnarray}\label{strict}
Tr \left( E^{Q_\mu}(A) E^{P_\mu}(B) \right)
< m_\mu(A) m_\mu(B).
\end{eqnarray}

In particular, the operators $Q_\mu$ and $P_\mu$ are not Accardi complementary
if $\mu > 0$.
\end{theorem}

{\bf Proof:}
We have shown the equality in (\ref{trace}), so
we only have to show that the integral is finite.
But this follows
since the integrand is continuous and the domain of integration is bounded.

We next claim that $ | \exp_\mu(ikx) | \le 1$ for $\mu > 0$ and that
this inequality is strict if $kx \ne 0$.
First for $\mu > 0$
note that $\exp_\mu(z) = \int_{-1}^1 \rd \eta_\mu (t) \re^{zt}$
for all $z \in \mathbb{C}$ by formula (2.3.5) in \cite{ros},
where $\rd \eta_\mu(t) = B(1/2,\mu)^{-1} (1-t)^{\mu -1} (1+t)^\mu \rd t$
is a probability measure on $[-1,1]$.
Here the normalization constant involves $B(1/2,\mu)$, a value of the beta function.
(See \cite{leb}.)
Then, it follows for all real $s \ne 0$ that
\begin{eqnarray}
 | \exp_\mu ( \ri s) |^2 = \left( \int_{-1}^1 \rd \eta_\mu (t)
\cos (st) \right)^2 +
\left( \int_{-1}^1 \rd \eta_\mu (t) \sin (st) \right)^2 \nonumber \\
\label{jens}
< \int_{-1}^1 \rd \eta_\mu (t) \cos^2 (st) + \int_{-1}^1 \rd \eta_\mu (t) \sin^2 (st)
= \int_{-1}^1 \rd \eta_\mu(t) =1,
\end{eqnarray}
where the inequality in an application of the strict form of Jensen's inequality,
given that the integrands are not constant, since $s \ne 0$.
(See \cite{ll} for Jensen.)
Clearly, $| \exp_\mu ( \ri s) | = 1$ if $s=0$.
But since the set $ (A \times B) \setminus \left(  \mathbb{R} \times \{ 0 \} \cup \{ 0 \}
\times \mathbb{R} \right)$
has positive $m_\mu \times m_\mu$ measure
and the set $ (A \times B) \cap \left(  \mathbb{R} \times \{ 0 \} \cup \{ 0 \} \times
\mathbb{R} \right)$
has zero $m_\mu \times m_\mu$ measure,
(\ref{strict}) now follows from (\ref{trace}) and (\ref{jens}).
QED.

Given that there are other inequalities in $\mu$-deformed analysis which hold in one
direction for $\mu > 0$ and
in the reverse direction when $-1/2 < \mu < 0$ and are equalities for $\mu=0$,
we conjecture that this holds here too,
namely, that
\begin{eqnarray}
\label{conj}
Tr \left( E^{Q_\mu}(A) E^{P_\mu}(B) \right)
> m_\mu(A) m_\mu(B)
\end{eqnarray}
for $A,B$ bounded Borel sets of positive $m_\mu$ measure and $-1/2 < \mu < 0$.
If this is conjecture is true,
then $Q_\mu$ and $P_\mu$ are not Accardi complementary
for $-1/2 <\mu < 0$.
Of course, Accardi showed the case of equality for $\mu=0$ in \cite{acc}.

We suppose that the technical hypothesis $ 0 \notin A^-$
in this theorem can be dropped without changing the result.

For the rest of this note we would like to discuss the possibility of getting
a more revealing formula for the integral in (\ref{trace}), for example something
that would help us prove conjecture (\ref{conj}).
Or can (\ref{trace}) be written in general
as the product $\nu_\mu(A) \nu_\mu(B)$ for {\it some}  measure $\nu_\mu$?
(This can be done, of course, for $\mu=0$.)

Therefore we wish to analyze the above integrand $ | \exp_\mu ( \ri kx ) |^2 $
in the general case $ \mu > -1/2$.
First we introduce the following definitions from Rosenblum \cite{ros}.
\begin{definition}
The {\em $\mu$-deformed binomial coefficient} is defined
for all non-negative integers $k$ and $j$ by
$ \left(
           \begin{array}{c}
                 k \\ j
           \end{array}
\right)_\mu := \frac{\gamma_\mu(k)}{\gamma_\mu(k-j)\gamma_\mu(j)}$.
The {\em $k$-th $\mu$-deformed binomial polynomial} is defined by
$ ~p_{k,\mu} (x,y) := \sum_{j=0}^k
\left(
           \begin{array}{c}
                 k \\ j
           \end{array}
\right)_\mu
x^j y^{k-j}$,
where $x,y \in \mathbb{C}$.
\end{definition}

Next we take $ s \in \mathbb{R}$ and find that
\begin{eqnarray*}
0 \le | \exp_\mu ( \ri s ) |^2 & = & \exp_\mu ( \ri s )  \exp_\mu ( - \ri s ) \\
& = & \sum_{l=0}^\infty \frac{1}{\gamma_\mu(l)} \ri^l s^l
\sum_{m=0}^\infty \frac{1}{\gamma_\mu(m)} (-\ri)^m s^m \\
& = & \sum_{k=0}^\infty \frac{1}{\gamma_\mu(k)} \ri^k s^k
\sum_{m=0}^k \frac{\gamma_\mu(k)}{\gamma_\mu(k-m)\gamma_\mu(m)} (-1)^m \\
& = & \sum_{k=0}^\infty \frac{1}{\gamma_\mu(k)} \ri^k s^k
\sum_{m=0}^k \left(
           \begin{array}{c}
                 k \\ m
           \end{array}
\right)_\mu
(-1)^m 1^{(k-m)} \\
& = & \sum_{k=0}^\infty \frac{1}{\gamma_\mu(k)} \ri^k s^k
p_{k,\mu} (-1,1) \\
& = & \sum_{j=0}^\infty \frac{1}{\gamma_\mu(2j)} \ri^{2j} s^{2j}
p_{2j,\mu} (-1,1) \\
& = & \sum_{j=0}^\infty \frac{ (-1)^{j} }{\gamma_\mu(2j)}
p_{2j,\mu} (-1,1) s^{2j}.
\end{eqnarray*}

We used here the identity $p_{k,\mu} (-1,1) = 0$ for $k$ odd.

Substituting this formula into the result for the trace we obtain
\begin{eqnarray*}
Tr \!\left( E^{Q_\mu}(A) E^{P_\mu}(B) \right)  \!=  \!\!\int_A \!\rd m_\mu
(x) \!\!\int_{ B } \!\rd m_\mu (k) \sum_{j=0}^\infty \frac{ (-1)^{j}
}{\gamma_\mu(2j)}
p_{2j,\mu} (-1,1) (kx)^{2j} \\
 =  \!\sum_{j=0}^\infty \frac{ (-1)^{j} }{\gamma_\mu(2j)}
p_{2j,\mu} (-1,1) \left( \int_A \rd m_\mu (x) x^{2j} \right)
\!\left( \int_{ B } \rd m_\mu (k) k^{2j} \right).
\end{eqnarray*}

We note the following formulas
for the $\mu$-deformed binomial polynomials:
\begin{eqnarray}
\nonumber
p_{0,\mu}(-1,1) &=&1 \\
\nonumber
p_{2n-1,\mu}(-1,1) &=& 0 \\
\nonumber
p_{4n-2,\mu} (-1,1) &=& \mu \, \frac{2^{2n-1} \prod_{k=n+1}^{2n-1} (\mu + k - 1 ) }{
\prod_{k=1}^{n} ( \mu + k - 1/2)  } \\
\nonumber
p_{4n,\mu} (-1,1) &=&
\mu \, \frac{2^{2n} \prod_{k=n+1}^{2n-1} (\mu + k ) }{ \prod_{k=1}^{n} ( \mu + k -
1/2)  } \\
\nonumber
p_{2n,\mu}(-1,1) &=& \frac{2 \mu}{n} \, \sum_{k=0}^{n-1} \left( \begin{array}{c} 2n
\\ 2k+1 \end{array} \right)_\mu
\end{eqnarray}

In all of these $n \ge 1$ is an integer.

The first two are readily proved, and the next three we have checked empirically
in a number of cases, and so we
believe them to be true.
However, we have not been able to use these to arrive at a more enlightening form of
the integral (and hence the trace) in formula (\ref{trace}).

\bigskip

\section[]{Conclusion}

\bigskip

As a concluding remark, we would like
to draw attention again to the conjectured inequality (\ref{conj}) and its
immediate consequence that $Q_\mu$ and $P_\mu$ are not
Accardi complementary for $-1/2 <\mu < 0$.

\bigskip

\section*{Acknowledgements}

\bigskip

The second author would like to thank Luigi Accardi for bringing to his
attention references \cite{acc} and \cite{cas-var} and thereby initiating
his interest in this fascinating topic during a visit in July, 2005
to the Centro Vito Volterra, Universit\`a di Roma ``Tor Vergata.''
He also thanks Luigi Accardi for his very warm hospitality.
This work was started during that visit.

\end{document}